\documentclass[fleqn,twoside,twocolumn,nofootinbib,showkeys]{revtex4} % Specifies the document class %,unsortedaddress
\usepackage[sec,nocpr]{ujp} % \usepackage[cyr]{ujp} for cyrillic
\begin{document}
\title[Multiplexing and Switching of Laser Beams]%колонтитул
{MULTIPLEXING AND SWITCHING\\ OF LASER BEAMS BASED ON CROSS-CORRELATION\\ INTERACTION OF PERIODIC FIELDS}%
\author{S.A.~Bugaychuk}%1
\affiliation{Institute of Physics, Nat. Acad. of Sci. of Ukraine}%институт 1
\address{46, Prosp. Nauky, Kyiv 03028, Ukraine}%адрес 1
\email{bugaich@iop.kiev.ua}%e-mail 1

\author{V.O.~Gnatovskyy}%2
\affiliation{Physical Department, Taras Shevchenko National University of Kyiv}%институт 2
\address{4, Prosp. Academician Glushkov, Kyiv 03022, Ukraine}%

\author{A.M.~Negriyko}%3
\affiliation{Institute of Physics, Nat. Acad. of Sci. of Ukraine}%институт 1
\address{46, Prosp. Nauky, Kyiv 03028, Ukraine}%адрес 1

\author{I.I.~Pryadko\,}%4
\affiliation{Institute of Physics, Nat. Acad. of Sci. of Ukraine}%институт 1
\address{46, Prosp. Nauky, Kyiv 03028, Ukraine}%адрес 1

\udk{535.4}  \razd{\seciv}

\autorcol{S.A.\hspace*{0.7mm}Bugaychuk, V.O.\hspace*{0.7mm}Gnatovskyy, A.M.\hspace*{0.7mm}Negriyko et al.}%

\setcounter{page}{301}%

\begin{abstract}
The  work is devoted to the study of the correlation technique of
formation of laser beams under conditions of the interaction of
coherent fields spatially periodic in the transverse
direction.\,\,It is aimed at the application of this technique to
the multiplexing (splitting) of an input laser beam to several
output ones, management of the energy of these beams, and their
clustering and debunching according to required time algorithms.
\end{abstract}


\maketitle

\section{Introduction}
The idea to use periodic diffraction elements for a directional
transformation of light beams was appeared simultaneously with the
invention of these elements.\,\,Well known is the importance of a
diffraction grating, which splits the incoming beam into several
secondary ones.\,\,In dependence on the structure of a groove of the
grating, one can amplify, decrease, or make equal the intensities in
the diffraction orders \cite{1,2}.\,\,Taking into consideration that
the autocorrelation function of any periodic distribution represents
also a periodic structure \cite{3}, the diffraction gratings are
applied to practically important correlation schemes for the
information processing (in particular, to the schemes with synthetic
aperture \cite{4}, holographic correlatora with discrete
representation of the signal \cite{5}, and so on).\,\,In the present
work, we are developing further capabilities to use periodic
structures for the management of laser beams, according to
\cite{6}.\,\,The purpose of the work is to increase the factors of
influence on the formed field and, consequently, to enlarge the area
of practical problems, which can be solved in such a way.\,\,We are
considering the task, when the interference pattern created by two
laser beams illuminates a flat phase diffraction grating, which can
be moved in the transverse direction.\,\,This problem represents a
special case of the correlation approach (two-stage method) to form
an optical field, which has a complex structure and a required
distribution of the energy at a technological target \cite{7}.\,\,In
addition, it can be considered as an extension of the dynamic
holography \cite{8,9} to the case where the phase elements, in which
a strong link between the distribution of the phase of the grating
and the intensity fringes of the interference pattern is broken, are
in use. In our method, the light field illuminating a phase mask has
a periodic structure.\,\,In so doing, both the phase distribution of
the field and the phase mask itself may be varied by changing their
periods, the structure of phase indices, and their mutual
shift.\,\,In particular, this distribution of the illumination
field, which is created by the interference of two or more beams,
can be replaced by its synthetic analog, which forms an
``equivalent'' field.\,\,The specially calculated synthetic
hologram, which is illuminated only by one beam, may be used as such
analog.\,\,In our work, we will investigate the ``equivalent'' field
formed by a periodic phase grating, which has the rectangular
profile.\,\,We will prove also that the field formed by such grating
substitutes a generally known pattern of interference
fringes.\,\,The lack of the obligatory correspondence of the
distributions of phase profiles between the input field and the
diffraction grating also promotes to introduce new dynamical
algorithms for the multiplexing, scanning, and switching of shaped
beams.

\section{Mathematical Background of the Correlation Method for Periodic Functions}

The basic scheme of the correlation method based on the periodic
structures for a transformation  of laser beams is shown in Fig.~1.

%Fig.~1
\begin{figure}
\includegraphics[width=\column]{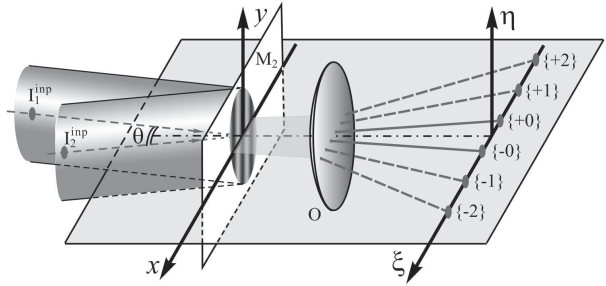}
\vskip-3mm\caption{Basic optical scheme of the correlation method
with two periodic structures.\,\,Two input plane waves
$I_{1}^{\mathrm{inp}}$ and $I_{2}^{\mathrm{inp}}$ converge at the
angle $\theta$ to form the interference pattern on the phase grating
$M_{2}$.\,\,The periods of the phase grating and the interference
fringes are either equal or multiple to one another.\,\,The
objective $O$ forms the angular spectrum of the resulting
field.\,\,The notations $\{\pm k\}$, $k=0,1,2,...$ mark the
diffraction orders on its output plane}
\end{figure}

According to this scheme, two plane waves form the interference
fringe pattern (let us call it the light grating), which illuminates
a periodic phase grating.\,\,The periodic intensity distribution on
the input is transformed into a set of diffraction orders at the
back focal plane of the objective $O$ on the output of the
system.\,\,The intensity of each diffraction order is easily
controllable by changing the parameters of the scheme, including the
mutual spatial shift of the periodic structures relative to one
another.

The main idea of the proposed technique consists in the fact that
one should create a ``coupled system'' of the light grating and the
material grating.\,\,In such coupled system, each diffraction order
on the output is formed as a result of the interference of partial
waves of the light grating.

The purpose becomes to find the special conditions to manage the
intensity in each diffraction order.\,\,That may be either the
amplification or suppression of some of them, nullification, or
doing them with equal intensities, what is made in the multiplexing
schemes.

The theoretical framework of the correlation technique with periodic
structures involves the known theorems for the displacement and for
the convolution of two functions.\,\,They are used to find matching
between the product of the distributions of two fields and the
convolution for the Fourier transforms (angular spectra) of these
fields.\,\,Thus, the correlation method of the formation of beams in
different diffraction orders lies in the consistent impact of
amplitude-phase converters on a previously diffracted field.\,\,In
the frame of this work, this is a transformation of the periodic
light field by means of a phase \mbox{grating.}\looseness=1

Now, let us consider the case where a plane light wave propagates
through two sequentially arranged modulators $M_1$ and
$M_2$.\,\,Note that the description of one of the modulators ($M_1$)
will correspond to the fringe pattern of the light intensity in the
interference area (see Fig.~1).

The angular spectrum of the output field is defined as the Fourier transform
of the diffraction field after these both modulators:
%1
\[
{\mathcal{\hat F}}\{M_1 M_2\}=\frac{1}{4\pi^2}\,{\mathcal{\hat F}}\{M_1\}\otimes
{\mathcal{\hat F}}\{M_2\}=
\]\vspace*{-6mm}
\begin{equation}
=\frac{1}{4\pi^2}\,m_1\otimes m_2,
\label{eq1}
\end{equation}
where $\mathcal{\hat F}$ and $\otimes$ are the operators of Fourier
transformation and convolution of functions,
respecti\-vely:\looseness=1
%2
\[
m_{1,2}\left(\xi,\eta\right)=\mathcal{\hat F}\{M_{1,2}\left(x,y\right)\}=
\]
\vspace*{-6mm}
\begin{equation}
=\int\limits_{-\infty}^{\infty}\!\int\limits_{-\infty}^{\infty}\!\!M_{1,2}\left(x,y\right)
\,\exp\!\left[2\pi i\left(\xi x+\eta y\right)\right]dx\,dy.
\label{eq2}
\end{equation}

\noindent Here, $\left(x,y\right)$ are the spatial coordinates  in
the plane of the modulators (in the front focal plane of the
objective, which makes the Fourier transformation of the light
field), $\left(\xi,\eta\right)$ are the angular coordinates in the
plane of the formation of diffraction orders (in the back focal
plane of the objective).

In the case of one-dimensional periodic structures, we can present the
transmittance of the modulators in the form of the convolution between the
distribution in a single ``window'' $S\left(x\right)$
(of an individual groove) and a comb function:
%3
\begin{equation}
\begin{array}{l}
M\left(x\right)=S\left(x\right)\otimes{\rm comb}\left(x\right)\!,
\\[1mm]
\displaystyle{\rm
comb}\left(x\right)=\sum\limits_{b=1}^{\infty}\delta\left(x-bT\right)\!,
\end{array}
\label{eq3}
\end{equation}
where $T$ is the width of the ``window'' $S\left(x\right)$.\,\,In
the extended form, we have the integral
%4
\begin{equation}
M\left(x\right)=\int\limits_{-\infty}^{\infty}\!\!S\left(x-y\right){\rm comb}\left(y\right)\,dy.
\label{eq4}
\end{equation}
\noindent
Thus, the output angular spectrum will be calculated according to
formula~(1) with regard for integral~(4),
which contains the specific profile of a converter.

\section{Calculations of Algorithms\\ for Switching and Spatial
Multiplexing\\
for the Correlation Scheme}

As an example, we consider the transformation of the interference
field with a sinusoidal intensity profile by means of a phase
diffraction grating, which has either rectangular (meander) or
triangular phase profile.\,\,Then we will have the following
expressions for the rectangular  and triangular phase profiles in
the ``window'', respectively:
%5
\begin{equation}
S_{\mathrm{rect}}\left(x\right)= -\Phi\,{\rm
sgn}\left(x\right)\!,~x\in\left[-T_{\mathrm{gr}}/2,T_{\mathrm{gr}}/2\,\right]\!,
\label{eq5}
\end{equation}\vspace*{-9mm}
%6
\begin{equation}
S_{\mathrm{tri}}\left(x\right)=
\Phi\!\left(1-2|x|/T_{\mathrm{gr}}\right)\!,~x\in\left[-T_{\mathrm{gr}}/2,T_{\mathrm{gr}}/2\,\right]\!,
\label{eq6}
\end{equation}
and\vspace*{-3mm}
%7
\begin{equation}
{\rm comb}\left(x\right)=\sum\limits_{b=1}^{\infty}\delta
\left(\!x-\frac{T_{\mathrm{gr}}}{2}\left(2b-1\right)+a\!\right)\!\!,
\label{eq7}
\end{equation}
where the constant $\Phi$ sets the phase profile of the
grating.\,\,We choose the physical size (the grating aperture) $L$
so that it includes an integer number $N$ of grooves, i.e.,
$L=NT_{\mathrm{gr}}$.\,\,The parameter $a$ determines the shift of
the grating along the coordinate $x$ and relative to the
interference fringes.

The input interference field is formed by two beams
$I_{1}^{\mathrm{inp}}$ and $I_{2}^{\mathrm{inp}}$ that have equal
intensities and plane wavefronts:
%8
\begin{equation}
\begin{array}{l}
{\mathbf{E}}_{1,2}={\mathrm{E}}\exp\left[i\left({
\mathbf{k}}_{1,2}\,{\mathbf{r}}- \omega
t+\varphi_{1,2}\right)\right]\!, \\[1mm]
\mathbf{E}=\left(0,E,0\right)\!,
\end{array}
 \label{eq8}
\end{equation}
where $\omega$ is the frequency of the laser radiation,
${\mathbf{k}}_{1,2}$ are the wa\-ve vec\-tors of the beams, and
$\varphi_{1,2}$ are their input phases.\,\,The wa\-ve vec\-tors have
the components
\[
{\mathbf{k}}_{1,2}=\left(\!\pm
k\sin\frac{\theta}{2},0,k\cos\frac{\theta}{2}\!\right)\!\!,
\]
and their interference occurs at the plane $(x,y)$.

Let, at the initial time moment, $\varphi_{1}=0$ and
$\varphi_{2}=\pi$. Then the expression for the interference pattern
has the form
%9
\begin{equation}
{\mathbf{E}}_{\mathrm{int}}\left(x\right)=2{\mathbf{E}}
\sin\left(\!k\sin\frac{\theta}{2}x\!\right)
\!\exp\left[-i\left(\omega t-\pi/2\right)\right]\!. \label{eq9}
\end{equation}
The angle of convergence of the beams $\theta$ is chosen from the
condition that $n$ numbers of the periods of the interference
pattern fits on the aperture $L$ of the diffraction grating. Then
formula (9) can be rewritten as follows:
%10
\begin{equation}
\begin{array}{l}
{\mathbf{E}}_{\mathrm{int}}\left(x\right)=2{\mathbf{E}}
\sin\left(2\pi nx/L\right) \exp\left[-i\left(\omega
t-\pi/2\right)\right]\!, \\[1mm]
x\in\left[0,L\right]\!.
\end{array}\!\!\!\!\!\!
\label{eq10}
\end{equation}
\noindent The period of the interference pattern
$T_{\mathrm{field}}$ will be related to the aperture of the grating
$L$ by the expression $T_{\mathrm{field}}=L/n$.\,\,The distribution
of the intensity of the interference pattern on the plane $(x,y)$ at
an arbitrary time in dependence on the coordinate $x$ will be
proportional to the expression
%11
\begin{equation}
I_{\mathrm{int}}\left(x\right)\propto \sin^2\!\left(2\pi
nx/L\right)\!,~x\in\left[0,L\right]\!. \label{eq11}
\end{equation}

The position and the intensities of diffraction orders are determined by
the following important parameters: by a relative spatial shift between the grating and
the interferences fringes, by the multiplicity of the period of the diffraction
grating relative to the interference pattern, and by a phase profile of the grating.

%Fig.~2
\begin{figure}
\vskip1mm
\includegraphics[width=\column]{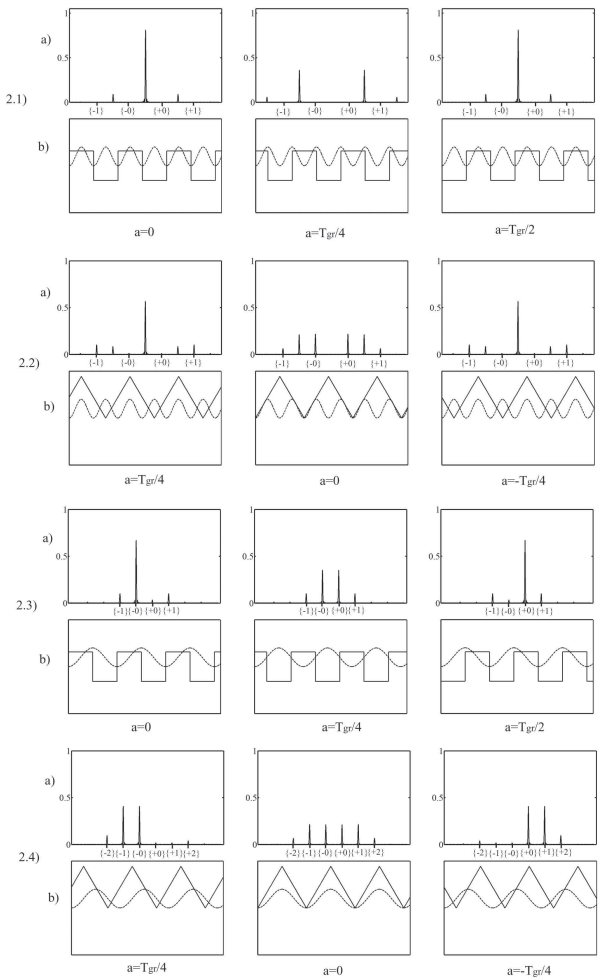}
\vskip-3mm\caption{Control over the intensities in the diffraction
orders under a shift of the diffraction grating.\,\,Positions
(\textit{a}) show the normalized intensities $I/I_{0}.$ Positions
(\textit{b}) show the relative spatial position of the phase profile
of the grating and the intensities in the interference pattern,
which are denoted by a dash curve.\,\,The notations $\{-0\}$ and
$\{+0\}$ designate the main diffraction orders, which correspond to
the positions of the beams, which form the interference
pattern.\,\,The first diffraction orders are marked by $\{-1\}$ and
$\{+1\}$.\,\,The value $a$ shows the spatial shift between the
maxima of the grating and the interference pattern.\,\,In
Figs.\,\,2.1 and 2.2, $n=20$ and $N=20$ are taken.\,\,In
Figs.\,\,2.3 and 2.4, these quantities are $n=10$ and $N=20$}
\end{figure}

We have found the conditions of the implementation of several
scenarios to make the spatial multiplexing, as well as the
management of  the intensities of the output beams, when the fringe
interference pattern illuminates a phase grating, which has either
the rectangular or triangular form of grooves.\,\,The results are
given in Fig.~2.

In Figs.~2.1 and 2.2, the period of the interference pattern is
doubled in comparison with the period of the diffraction
grating.\,\,The central peak in Fig.~2.1, which is created under the
conditions of $a=0$ and  $a=T_{\mathrm{gr}}/2$, may be redistributed
to two channels (left and right) with equal intensities under the
condition $a=T_{\mathrm{gr}}/4$.\,\,Therefore, the parameter of the
switching of channels is the mutual spatial shift $a$ between the
light periodic structure and the diffraction
\mbox{grating.}\looseness=1

If we replace the rectangular grating by a triangular one, then the
left and right channels \mbox{will} become doublets, as it is seen
from Fig.\,\,2.2 under the condition $a=0$.\,\,Thus, one can create
four beams with the same intensity.\,\,We have obtained the
possibility to form either one, or two, or four beams of equal
intensities by changing the grating profile and the mutual
transverse shift.\,\,Owing to the simple structures of the phase
profiles, the intensities of created beams can be easily adjusted
with the use of computer-controlled phase mo\-dulators.\looseness=1

In Figs.~2.3 and 2.4, the periods of the light pattern and of the
diffraction grating are the same. For the rectangular grating
profile, one can create two channels for the switching: we can
either leave only the left one (Fig.\,\,2.3 for $a=0$), or only the
right one ($a=T_{\mathrm{gr}}/2$), or the both simultaneously
($a=T_{\mathrm{gr}}/4$).\,\,In the case of a triangular grating
profile (Fig.\,\,2.4) under the condition $a=0$, four diffraction
maxima (communication channels) with equal intensities are
created.\,\,If the shift is changed by the value of
$a=T_{\mathrm{gr}}/4$, only two left channels appear, and there are
only two right channels for the shift
\mbox{$a=-T_{\mathrm{gr}}/4$.}\looseness=1

Each of the described scenarios for the change of both the intensity
and the location of the diffraction orders are determined due to the
shift, phase relief, and mutual dispositions of the periodic
structures.\,\,Therefore, all these factors are important to make
manipulation of the spatial channels on the output of the
correlation scheme.\,\,On a practical level, controlling these
parameter may be implemented with the help of a spatial
electro-optical modulator of phases, which is placed on the area of
the formation of interference (or created by another way) fringes.
This modulator could create a grating, parameters of which can be
changed with a \mbox{computer.}\looseness=1

\section{Experimental Study}\vspace*{-1.5mm}

We have studied the angular spectra of the generated fields, namely
the distribution of the intensities in the diffraction orders of the
correlation scheme.\,\,The optical scheme of the experimental set-up
corresponds to Fig.~2.1.\,\,The radiation of a He--Ne laser was
expanded by a telescopic system and filtered by a system of
diaphragms.\,\,The resulting beam with uniform distribution of the
energy in the cross-section and with a plane wavefront is forwarded
onto a beam-splitter device to obtain two beams for their later
interference.\,\,Moreover, the block of the beam-splitter device
provides a fine adjustment of the angle between two beams.\,\,These
two beams form the interference pattern, which illuminates a
diffraction grating.\,\,The grating is mounted in a holder, which
adjusts a required orientation of the grating in the horizontal and
in the vertical directions.\,\,The necessary ratio between the
period of interference fringes and the period of the diffraction
grating is achieved by the adjustment of the convergence angle
between the interfering \mbox{beams.}\looseness=1

The angular spectrum of the diffraction field is observed on a back
focal plane of a high-quality Fourier objective.\,\,The beams of the
diffraction orders, after a necessary magnification, have been
directed to photodiodes by a system of prisms.\,\,Their oscillograms
have been registered and then processed by a computer.\,\,In the
experimental investigations, we used the phase gratings, which had a
rectangular profile.

%Fig.~3
\begin{figure}
\vskip1mm
\includegraphics[width=\column]{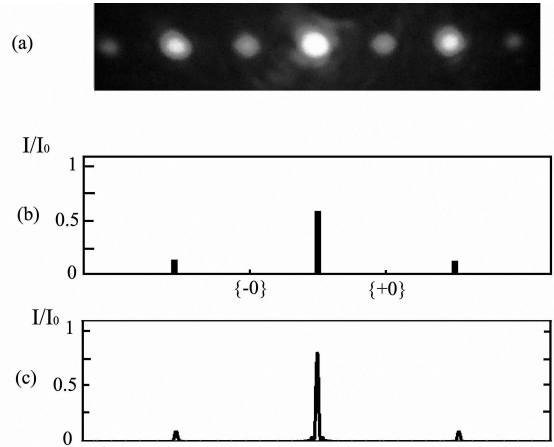}
\vskip-3mm\caption{Experimental measurements of the redistribution
of the intensities in the diffraction orders for a rectangular
grating with double period relative to the interference fringes (see
case 2.1 in Fig.\,\,2), and their comparison with theoretical
calculations: the photos of the diffraction orders (\textit{a});
histograms of the measured intensities (\textit{b}); calculated
intensities (\textit{c}).\,\,The notations $\{-0\}$ and $\{+0\}$
denote the angular spectra of the beams, which form the interference
pattern.\,\,The mutual shift is $a=0$ }
\end{figure}
%Fig.~4
\begin{figure}
\vskip5mm
\includegraphics[width=\column]{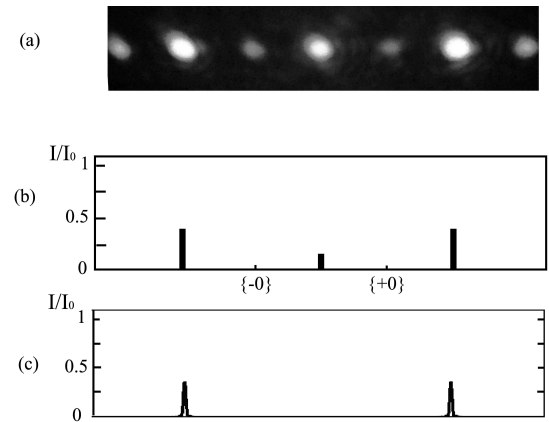}
\vskip-3mm\caption{The same as in Fig.\,\,3.\,\,The mutual shift is
$a=T_{\mathrm{gr}}/4$}
\end{figure}

%Fig.~5
\begin{figure}
\vskip1mm
\includegraphics[width=7.5cm]{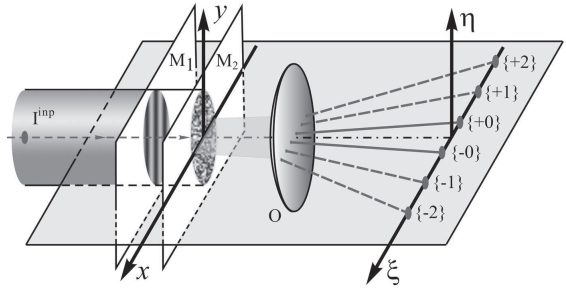}
\vskip-3mm\caption{Basic optical scheme for the correlation
technique with periodic structures based on the ``equivalent''
interference pattern.\,\,The input plane wave is $I^{\mathrm{inp}}$,
the equivalent phase mask is $M_{1}$, i.e.\,\,this is an additional
phase grating with determined phase profile.\,\,The diffraction
grating and the equivalent mask have either equal periods or
multiple periods.\,\,The other designations are the same as in
Fig.~1}
\end{figure}
%Fig.~6
\begin{figure}[h!]
\vskip1mm
\includegraphics[width=\column]{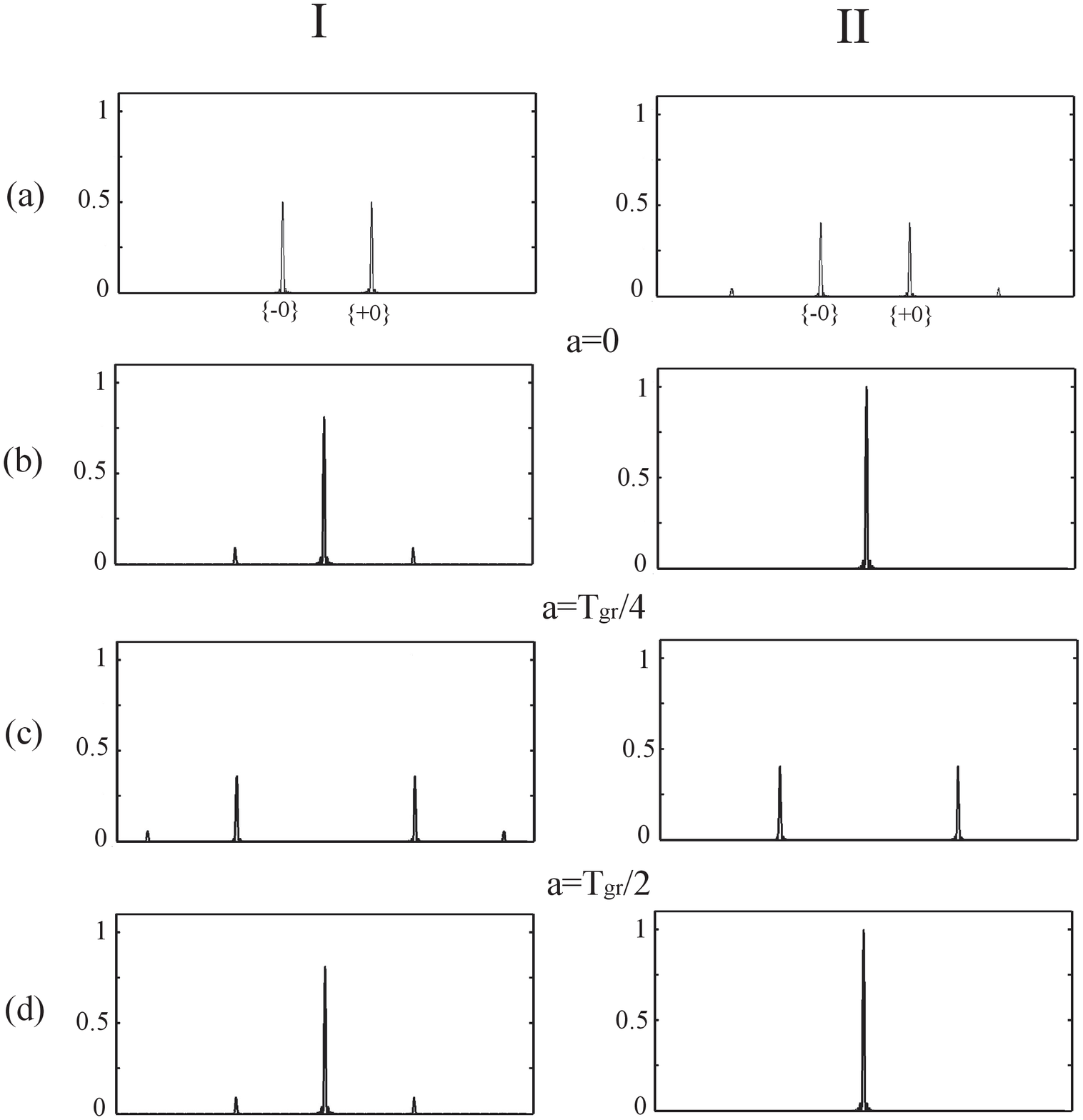}
\vskip-3mm\caption{Comparative calculations of the intensities in
the diffraction orders for the scheme with ``real'' interference
pattern I (scheme in Fig.~1) and for the scheme with ``equivalent''
pattern II (scheme in Fig.~5).\,\,The phase diffraction grating
$M_{2}$ has the rectangular profile of a relief shown in Fig.~2.1.
Figure (a) shows the main diffraction orders on the output plane of
the scheme in the case where the diffraction grating $M_{2}$ is
absent.\,\,The value $a$ determines the spatial shift between the
periodic structures}%\vspace*{-2mm}
\end{figure}
%Fig.~7
\begin{figure}
\vskip1mm
\includegraphics[width=\column]{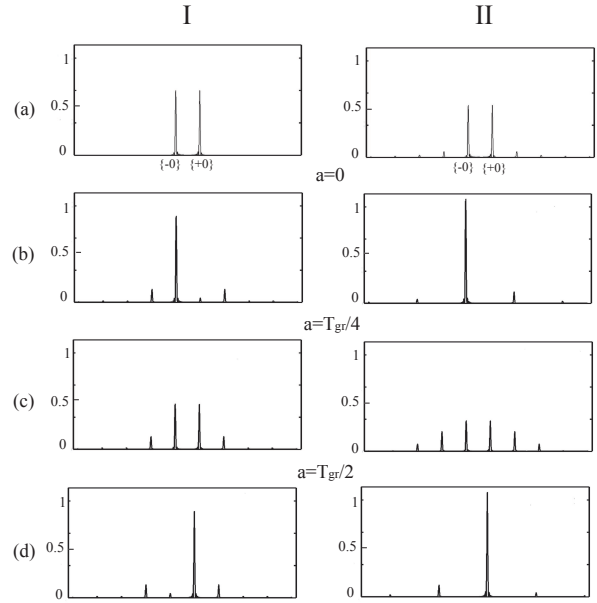}
\vskip-3mm\caption{Comparative calculations of the intensities in
the diffraction orders for the scheme with ``real'' interference
pattern I (scheme in Fig.~1) and for the scheme with ``equivalent''
pattern II (scheme in Fig.~5).\,\,The phase diffraction grating
$M_{2}$ has the rectangular profile of the relief shown in Fig.~2.3.
Figure (\textit{a}) shows the main diffraction orders on the output
plane of the scheme in the case where the diffraction grating
$M_{2}$ is absent.\,\,The value $a$ determines the spatial shift
between the periodic structures}\vskip2mm
\end{figure}

The results of experimental researches of the management of the
angular spectrum and their comparison with the theoretical
calculations are presented in Fig.~3.\,\,The grating has the
rectangular phase profile with the modulation depth equal to
$\pi/2$.\,\,Such value of the modulation depth can be easily
controlled experimentally, because, in this case, one can observe
the appearance of a strong maximum between $\{-0\}$ and $\{+0\}$
diffraction orders together with a strong decrease in the
intensities in these main diffraction orders.\,\,This effect is well
observed in Fig.~3.\,\,In the experiments, we apply a usual
interference pattern  on the input of the system that was formed by
two plane waves of equal intensities.\,\,In Figs.~3 and 4, the data
of the experimental implementation of two scenarios depending on the
spatial shift of the diffraction grating relative to the
interference pattern and the comparison with the theoretical
calculations are given.

Moreover, the measured values of the normalized intensities
practically  coincide with the theoretical calculations: in Fig.~3,
the measured values are ${I/I_{0}}^{\exp}=0.6$ for the central
maximum and ${I/I_{0}}^{\exp}=0.14$ for the side peaks, whereas
corresponding calculated values are ${I/I_{0}}^{\mathrm{theor}}=0.8$
and ${I/I_{0}}^{\mathrm{theor}}=0.12$.\,\,In the Fig.~4 the measured
intensities are ${I/I_{0}}^{\exp}=0.4$ for the side peaks and
${I/I_{0}}^{\exp}=0.15$ for the central peak, the calculated
intensities for the side peaks are
${I/I_{0}}^{\mathrm{theor}}=0.37$.\,\,Note, that the central maximum
is damped not completely in the experiments.\,\,We explain it by the
fact that the two input waves could not be entirely equal in a real
experimental \mbox{set-up.}%\looseness=1

\section{Equivalent Mask with one Plane Wave\\ for the Substitution of Interference Pattern}

For practical using of the correlation scheme, there will be
appropriate the exclude the interferometer, which play a role of a
``generator'' of a light periodic pattern in a form of interference
fringes.\,\,The best decision for this purpose is the substitution
of the interferometer (as well as the interference pattern) by some
amplitude-phase mask, which has such the complex amplitude of the
transmittance that is identical to the complex amplitude describing
the interference pattern, and therefore, it has the following form:
%12
\[
M_{\mathrm{equ}}\left(x\right)=|\sin\left(2\pi
x/T_{\mathrm{field}}\right)|\times
\]\vspace*{-7mm}
\begin{equation}
\times\,\exp\left[i\pi/2\,{\rm sgn}\left(\sin\left(2\pi
x/T_{\mathrm{field}}\right)\right)\right]\!. \label{eq12}
\end{equation}

After the illumination of this mask $M_{\mathrm{equ}}\left(x\right)$
with the help of one flat wave, we obtain the same distribution of
the fringe pattern, like in the previous cases, on the
plane~$\left(x,y\right)$, where the diffraction grating is
set.\,\,The calculations with the use of this mask give, naturally,
the same results, as in the case of using the interference
pattern.\,\,For the only exception, that the amplitude of
distribution (12) takes a half of the energy of the beam
illuminating this mask.\,\,Therefore, it has the sense to consider
the case in the correlation technique where one uses a distribution
similar to (12), but another mask, which contains only the
phase part of $M_{\mathrm{equ}}\left(x\right)$, i.e,
\begin{equation}
M_{\mathrm{equ}}\left(x\right)= \exp\left[i\pi/2\,{\rm
sgn}\left(\sin\left(2\pi
x/T_{\mathrm{field}}\right)\right)\right]\!. \label{eq13}
\end{equation}
One can see that such mask represents a simple periodic phase
structure with rectangular profile of a relief with the depth
equals to $\pi$ radians, in fact, the diffraction grating with
rectangular profile of the grooves. Such grating can be easily
fabricated or be simulated on the controllable phase
mo\-dulator.\looseness=1

The experimental set-up of the correlation scheme with equivalent
mask is shown in Fig.~5, and it is differ only by the input part
comparing with the scheme in Fig.~1.\,\,In the case of the
``artificial'' interference pattern, only one plane wave comes on
the input of the optical scheme, and it goes through two diffraction
gratings, which are sequentially arranged close to each
other.\,\,The phase distribution in the equivalent grating is
rectangular according to expression (13), but the phase
distribution for the grating of a converter may be arbitrary, in
particular, like it was described in the previous
\mbox{section.}\looseness=1

In the Figs.~6 and 7, we present the results of calculations
corresponding to the use of the ``equivalent'' mask (13)
together with the converter grating with the rectangular profile of
a groove.\,\,In Fig.~6, the diffraction grating of a converter has
the rectangular phase profile with the modulation depth $\pm\pi/2$
and the double period compared with the period of the equivalent
mask (see Fig.\,\,2.1).\,\,In Fig.~7, the converter has the
rectangular phase profile with the modulation depth $\pm\pi/4$ and
the same period as the period of the equivalent mask (see
Fig.~2.3).\looseness=1

One can see that the substitution of the interference pattern by
the equivalent mask gives very similar results, and they are
completely suitable in practical schemes of the switching and
mul\-tiplexing.\looseness=1

We have carried out the experimental researches, which are similar
to those described in Section~4.\,\,The same diffraction converters
$M_{2}$ are used in this experiments, but, instead of the
interference pattern, we apply the equivalent phase mask $M_{1}$
placed before the converter $M_{2}$.\,\,So that, the plane wave
firstly diffracts on the equivalent mask.\,\,After that, the created
field diffracts on the grating.\,\,The latter is set in the near
field of the mask $M_{1}$ at the distance $\le$$ 200\lambda,$
according to our estimations.\,\,We have obtained that the
equivalent mask very efficiently replaces the interference pattern.

\section{Conclusion}

The application of the correlation technique based on a periodic
structure expands capabilities of the schemes for the multiplexing
of a laser beam for such problems, where one needs changes of the
intensities in the output beams, their location on a technological
target, as well as for the switching of the optical
channels.\,\,During this process, one can implement dynamic
scenarios of the multiplexing, that are almost not attainable during
the traditional use of one diffraction  grating even with a
complicated profile of an individual groove.\,\,These advantages of
the correlation technique are achieved due to increasing the number
of factors, which impact the formation of beams.\,\,In particular,
they are the dynamics of a transverse shift of two gratings relative
to each other and the possibility to change the multiplicity of
their periods.\,\,At the same time, the correlation approach
conserves a relative simplicity of a phase relief in individual
grooves in both used modulators, which facilitates their practical
application as computer-controllable \mbox{converters.}\looseness=1

\rezume{%Резюме укр. мовою
С.А.\,Бугайчук, В.О.\,Гнатовський,\\ А.М.\,Негрійко, І.І.\,Прядко}
{МУЛЬТИПЛІКАЦІЯ ТА КОМУТАЦІЯ\\ ЛАЗЕРНИХ ПУЧКІВ ПРИ
КРОС-КОРЕЛЯЦІЙНІЙ\\ ВЗАЄМОДІЇ ПЕРІОДИЧНИХ ПОЛІВ} {Робота присвячена
дослідженню кореляційного методу формування лазерних пучків під час
взаємодії періодичних у поперечному напрямку когерентних полів і
спрямованого на застосування цієї методики для мультиплікації
(розщеплення) первинного лазерного пучка на декілька вторинних,
керуванням величиною енергії в цих пучках, їх групуванням і
розгрупуванням згідно з потрібним часовим алгоритмом.}

\end{document}